\documentclass[12pt]{iopart}
\usepackage{graphicx}

\begin{document}

\title[Decoherence and robustness...]{Decoherence and robustness of parity dependent
entanglement in the dynamics of a trapped ion}
\author{ S. Maniscalco, A. Messina, A. Napoli}

\address{INFM Unit\`{a} di Palermo and  Dipartimento di Scienze
Fisiche ed Astronomiche dell'Universit\`{a} di Palermo, via Archirafi
36, 90123 Palermo, Italy  }

\author{D. Vitali}

\address{INFM Unit\`{a} di Camerino and Dipartimento di Matematica
e Fisica dell'Universit\`{a} di Camerino, via Madonna delle Carceri,
62032 Camerino, Italy  }

\begin{abstract}
We study the entanglement between the 2D vibrational motion
and two ground state hyperfine levels of a trapped ion, Under
particular conditions this entanglement depends on the parity of the
total initial vibrational quanta. We study the robustness
of this quantum coherence effect with respect to the presence
of non-dissipative sources of decoherence, and of an imperfect
initial state preparation.

\end{abstract}

\section{Introduction}
Over the last few years we have witnessed a very rapid development of
cooling and trapping techniques both for neutral atoms and for ions
\cite{itano82,winel87,died89,monr95,ref1,ref2,ref3}. These progresses have made it
possible to perform sophisticated experiments wherein several
examples of couplings between a few bosonic and fermionic dynamical
variables have been realised
\cite{monr952,monr95,meek96,monr96,blatt,schmidt}. If, in fact, an
ion confined in a miniaturised Paul trap is exposed to suitably
configured laser beams, the 3D harmonic motion of the ionic center of
mass gets entangled with the internal degrees of freedom. Some
peculiar aspects of the vibronic response stemming from such a
dissipation-free situation have been successfully exploited
for  realizing experimentally Fock states, coherent states, squeeezed states and Schr\"{o}dinger cat-like states \cite{meek96,monr96,blatt},  presenting theoretical schemes for engineering  several nonclassical state
\cite{cirac93,cirac94,ciracb,blatt85,zeng95,
gerry97b,gouc96,goua96,goub96,sab,sab2},
implementing  quantum logic gates \cite{monr952,ci,li}, realising
tomographic reconstructions of the density matrix of the
system \cite{lei,wal,poy,bard}, and, more in general,  discovering
quantum effects characterising the ionic quantized oscillatory
motion \cite{wa97,hu99,wa297,wa99,ma98,st97}.
Many hamiltonian models have been reported so far in the literature to
describe physical properties of trapped ions. Some models explore
physical situations wherein only the ionic motion along a specified
direction of the trap is effectively influenced by the presence of
the laser beams. When the induced vibronic coupling is instead
extended over two (three) independent directions of the trap, then
one refers to the correspondent physical scenario as to a 2D (3D)
trapped ion.

In this paper we study the motion of an ion isotropically confined in
the radial plane of a  Paul microtrap  when it is irradiated by a
properly chosen configuration of external laser beams. We show that
there exist experimentally interesting conditions under which the
dynamics of this system may be exactly treated \cite{sab}. Exploiting
this fact we
find that the center of
mass vibrational motion and the electronic degrees of freedom of the
ion becomes entangled, in a way which is very sensitive to the
parity of the initial number of vibrational quanta. Then we study
the robustness of this quantum effect with respect to various
experimental imperfections. In fact, even if dissipation of the
vibrational motion is completely negligible, non-dissipative sources
of decoherence associated with fluctuations of external parameters
may be important. We study these effects, and those of laser intensity
fluctuations in particular, using the
model-independent formalism of Ref.~\cite{bonifacio}. We find that,
using standard values for the strength of the laser intensity
fluctuations, the parity-dependent entanglement effect
is not washed out by non-dissipative decoherence.
The effect of an imperfect initial state preparation of the
vibrational degree of freedom is also studied.

\section{Parity dependent entanglement effect}

\subsection{Derivation of the effective Hamiltonian model}
Consider a two-level ion of mass $M$ confined in a bidimensional isotropic harmonic potential characterised by
the trap frequency $\nu$. Let's denote by $\hbar \omega_0$ the energy separation between the ionic excited state
$\vert + \rangle$ and its ground state $\vert - \rangle$ and assume $\omega_0\gg \nu$. Indicate by $a$
($a^{\dag}$) and $b$ ($b^{\dag}$) the annihilation (creation) operators of vibrational quanta relative to the
ionic center of mass oscillatory motion along the $x$ and $y$ axes of the radial plane of the trap respectively.

We assume that the ion is driven by two $\pi$-out of phase laser beams, applied along the two orthogonal
directions $\bar{x}$ and $\bar{y}$  with an angle of $\pi / 4$  relative to the $x$ and $y$ axis respectively.
The two lasers have equal frequency  $\omega_L=\omega_0-2\nu$, intensity $E_0$ and wavevector modulus $k_L$.
After performing the dipole and the rotating wave approximations, the  hamiltonian model of the system can be
cast in the form
\begin{equation}
\hat{H} =\hbar \nu (\hat{a}^{\dag} \hat{a} + \hat{b}^{\dag} \hat{b} )+ \frac{\hbar \omega_0}{2} \hat{\sigma}_z +
[d\epsilon^{(-)} (\bar{x}, \bar{y}, t) \hat{\sigma}_- + h.c.]
\end{equation}
where $d$ is the appropriate dipole matrix element, assumed real for simplicity,  and
$\epsilon^{(-)}(\bar{x},\bar{y},t)= E_0 [e^{i(\omega_L t-k_L \bar{x})}-e^{i(\omega_L t-k_L \bar{y})}]$ represents
the negative frequency contribution to the resultant electric field acting upon the atom. The two-level atomic
degrees of freedom are described in terms of the inversion operator  $
 \hat{\sigma}_z=\vert + \rangle \langle
+ \vert -\vert - \rangle \langle - \vert $ and  of the two transition operators $ \hat{\sigma}_{\pm}=\vert \pm
\rangle \langle \mp \vert$. In a reference frame rotating at the laser frequency $\omega_L$,the hamiltonian
$\hat{H}$ assumes the form $\hat{H}_L=\hat{H}_0+\hat{H}_1$ with
\begin{equation}
\hat{H}_0=\hbar \nu (\hat{a}^{\dag} \hat{a} + \hat{b}^{\dag} \hat{b} )+ \frac{\hbar \delta}{2} \hat{\sigma}_z
\end{equation}

\begin{equation}
\hat{H}_1=\hbar \Omega \{[e^{-i \eta(\hat{A}+\hat{A}^{\dag})}-
e^{-i\eta(\hat{B}+\hat{B}^{\dag})}]\hat{\sigma}_-+h.c.\}
\end{equation}
where
\begin{equation}
\delta= \omega_0-\omega_L=2\nu,\;\;\;\;\; \Omega =\frac{dE_0}{\hbar},\;\;\;\;\; \eta=k_L\sqrt{\frac{\hbar}{2 \nu
M}}
\end{equation}
are the atom-laser detuning, the Rabi frequency and the so-called Lamb-Dicke parameter respectively. We have
indicated with $\hat{A}$ ($\hat{A}^{\dag}$ ) and  $\hat{B}$ ($\hat{B}^{\dag}$ ) the annihilation (creation)
operators of vibrational quanta along the directions $\bar{x}$ and $\bar{y}$ respectively, defined by
\begin{equation}
\hat{A}=\frac{1}{\sqrt{2}} \left( \hat{a} + \hat{b} \right) \hspace{1cm} 
\hat{B}=\frac{1}{\sqrt{2}} \left( \hat{b} - \hat{a} \right) 
\label{AeB}
\end{equation}
In the interaction picture, the interaction Hamiltonian assumes the form
\begin{eqnarray}
\hat{H}_{int} &=&  \hbar \Omega e^{-\eta ^2 /2} \left\{ \sum_{k,j=0}^{\infty} (-1) ^{j+k} \frac{(i
\eta)^{j+k}}{j!k!} \hat{A}^j \left( \hat{A}^{\dag} \right) ^k e^{i  \nu t (k-j-2)} + \right. \nonumber \\
 &-& \left. \sum_{k,j=0}^{\infty}  (-1) ^{j+k} \frac{(i \eta)^{j+k}}{j!k!} \hat{B}^j \left(
 \hat{B}^{\dag}\right)^k  e^{i  \nu t (k-j-2)} \sigma_-+ h.c. \right\} \label{hint1}
 \end{eqnarray}
where the resonance condition $\delta = 2 \nu$ has been used. 
Under the assumption
 $\nu \gg \Omega$, we may perform the rotating wave approximation neglecting in Eq. (\ref{hint1}) all the terms
 oscillating at multiples of the harmonic trap frequency $\nu$. Finally, using Eqs. (\ref{AeB}) it is not
 difficult to show that, in the Lamb-Dicke limit $\eta   \ll 1$, the effective interaction
 Hamiltonian in the interaction picture can be written as \cite{gerry97b,goua96}
 \begin{equation} \hat{H}_{int}= \hbar g \left(
 \hat{a} \hat{b}\; \hat{\sigma}_+ + h.c. \right) \label{hamil}
 \end{equation}
 with $g = \Omega \eta^2 e^{-\eta^2/2}$.

\subsection{Experimental set up}
For realizing the Hamiltonian model (\ref{hamil}) we make use of a three level electronic system as shown in
Fig.~1. Such a system is similar to the one currently used in the experiments 
performed at NIST \cite{meek96,monr96}. The weak transition $\vert - \rangle \leftrightarrow \vert + \rangle$ is
simultaneously driven by two Raman laser configurations with different detunings ($\Delta_1$ and $\Delta_2$) from
the electronic level $\vert 3 \rangle$ and with frequency differences $\omega_L= \omega_{L_1}-\omega_{L_2} \equiv
\omega_{L'}= \omega_{L'_1}-\omega_{L'_2} = \omega_0 - 2 \nu $. For sufficiently large detunings,  such that the
intermediate state $\vert 3 \rangle$ can be adiabatically eliminated, only the electronic levels $\vert +
\rangle$ and $\vert - \rangle$ are involved in the dynamics.  In this case the two-photon Raman transitions
between the states of interest are formally equivalent to narrow single photon transitions \cite{meek96,monr96}.
The Raman laser fields are chosen to have wave vector differences $\overrightarrow{k}_L= \overrightarrow{k}_{L_1}
- \overrightarrow{k}_{L_2} $ and $\overrightarrow{k}_{L'}= \overrightarrow{k}_{L'_1} - \overrightarrow{k}_{L'_2}
$ pointing in the $\bar{x}$ and $\bar{y}$ directions respectively. Under these conditions, in the resolved
sideband regime and in the Lamb-Dicke limit, the dynamics of the system is described by the Hamiltonian
(\ref{hamil}) provided that we make the following substitution $g=\Omega_R \eta_R^2 e^{-\eta_R^2/2}$ with
 $ \Omega_{R}= \Omega_{L_1}\Omega_{L_2} / \Delta_1  = \Omega_{L'_1} \Omega_{L'_2} / \Delta_2$ and $\eta_{R} =
\sqrt{\hbar k_L^2 / 2 M \nu}  = \sqrt{\hbar k_{L'}^2 / 2 M \nu}$. The dependence of the Lamb-Dicke parameter on
the modulus of the wave vector difference allows to vary $\eta_R$ by changing the Raman beams configurations. In
this way it is possible to obtain very small values of the Lamb-Dicke parameter as required in our scheme.

For the scope of the paper it is important to underline that measurements of the electronic states are
experimentally realizable with very high efficiency by means of the so called quantum jumps tecnhique
\cite{meek96,monr96,blatt}. The method essentially consists in coupling the ground electronic level $\vert -
\rangle$ to a third auxiliary   level $\vert r \rangle$ by means of another laser beam. The transition $\vert -
\rangle \leftrightarrow \vert r \rangle$ is chosen to be a dipole allowed one. In this conditions, the presence
of fluorescence detects the ion in the electronic ground level and its absence 
in the excited level. With this method state detection efficiency
approaches $100 \%$, even if a
modest number of scattered photons are detected.

\begin{figure}
\centering
 \includegraphics[width=8cm]{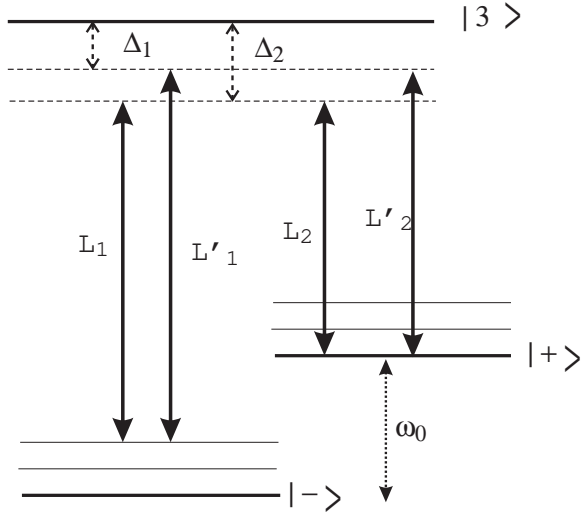}
\caption{Three-level electronic system of the trapped ion and 
scheme of Raman couplings}
\end{figure}

\subsection{Dynamics}
 Let's denote with $\vert n_a, \: n_b \rangle = \vert n_a \rangle \vert
n_b \rangle$ the product of the vibrational Fock states along the $x$ and $y$ directions and suppose that the
initial state of the ion has the form
\begin{eqnarray}
\vert \Psi (0) \rangle   =\vert \tau = 1, j_0= \frac{N}{2} \rangle
\vert - \rangle&\equiv& \frac{1}{2^{N/2}} \sum_{k=0}^{N} \left(
\begin{array}{c} N \\ k \end{array} \right)^{1/2}  \vert N-k,k
\rangle  \vert - \rangle\nonumber \\
&\equiv& \sum_{k=0}^{N} P_k  \vert N-k,k \rangle  \vert - \rangle
\label{su2}
\end{eqnarray}
The vibrational state $\vert \tau = 1, j_0= \frac{N}{2} \rangle $
belongs to the class of the so called SU(2) coherent states defined as
\begin{equation}
\begin{array}{ll}
\vert \tau , j \rangle= \frac{1}{\left( 1+|\tau|^2 \right) ^j}
\sum_{k=0}^{2j} \left( \begin{array}{c} 2j \\ k \end{array}
\right)^{1/2} \tau ^k  \vert 2j-k,k \rangle    \label{su2gen}
\end{array}
\end{equation}
where $\tau \in C, \: 2j \in N $. The states $\vert N-k, k \rangle$ appearing in Eq.\ (\ref{su2})  are
eigenstates of the operator $\left( \hat{a}^{\dag} \hat{a}+\hat{b}^{\dag} \hat{b} \right)$ all pertaining to the
eigenvalue $N=2j_0$ representing the initial total number of vibrational quanta. In particular, as pointed out by
Gou and Knight \cite{goub96}, the SU(2) coherent state corresponding to the specific value $\tau=1$, $\vert \tau
=1, j_0 \rangle$ of a bidimensionally confined ion, coincides with the vibrational Fock state with $N=2j_0$
quanta along the direction $\bar{x}$ forming an angle $\pi / 4$ relative to the $x$ axis. Since number states for
the 1D motion of an ion in a rf-trap have been already experimentally realized (see \cite{meek96,blatt}), it is
evident that the initial state of Eq.~(\ref{su2}) could be easily prepared experimentally.

Then, if at $t=0$ the laser fields realizing the coupling
between the 2D vibrational motion and the internal two-level system
described by the Hamiltonian
model of Eq.~\ (\ref{hamil}) is turned on, at any subsequent time instant
$t$, the state of the system can be written as
\begin{equation}
 \vert \Psi(t) \rangle = \vert \varphi _- (t) \rangle \vert - \rangle
+ \vert \varphi _+ (t) \rangle \vert + \rangle \label{state}
\end{equation}
with
\begin{equation}
\vert \varphi _- (t) \rangle = \sum_{k=0}^{N}P_k \cos(f_k t)\vert
N-k,k \rangle \label{fi-}
\end{equation}
and
\begin{equation}
\vert \varphi _+ (t) \rangle = -i\sum_{k=1}^{N-1}P_k \sin(f_k t)\vert
N-k-1,k-1 \rangle \label{fi+}
\end{equation}
where
\begin{equation}
f_k = 2g \sqrt{(N-k)k} \label{frabi}
\end{equation}
are the Rabi frequencies.
Eq.\ (\ref{state}) shows that, starting from the factorized initial state
$\vert \Psi(0) \rangle $ of Eq.~(\ref{su2}), the Hamiltonian model
(\ref{hamil})
generates entanglement between the external and internal degrees of
freedom of the trapped ion, giving rise to interesting
dynamical consequences.
In order to appreciate the meaning of this assertion, we focus our
attention on the time evolution of the vibrational entropy
\begin{equation}
S_v(t)=-Tr[ \rho _v(t) \: \ln \: \rho _v(t)] \label{entr1}\;,
\end{equation}
$\rho_v$ being the reduced density operator describing the 2D
vibrational motion of the ion. A straightforward calculation gives
\begin{equation}
S_v (t)= -\ln \left[ c(t)^{c(t)} [1-c(t)]^{1-c(t)} \right]
\label{entr2}
\end{equation}
where
\begin{equation}
c(t)= \sum_{k=0}^N \left| P_k \right| ^2 \cos^2 (f_k t)=  \frac{1}{2}
\left[ 1+ \sum_{k=0}^N  \left| P_k \right| ^2 \cos (2f_k t)\right] \;.
\label{ci}
\end{equation}
\begin{figure}
\centering
 \includegraphics[width=8cm]{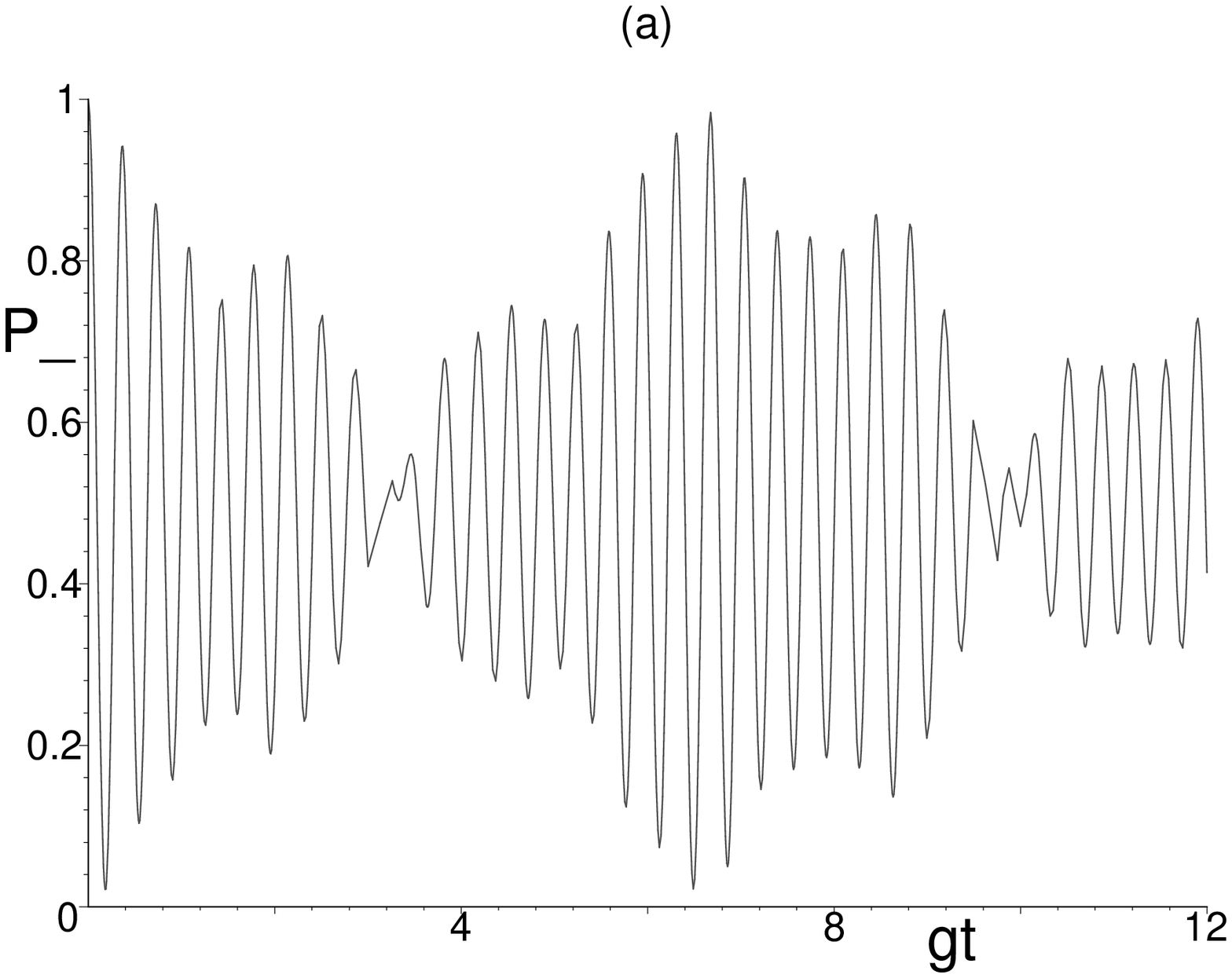}

\vspace{0.5cm}

 \includegraphics[width=8cm]{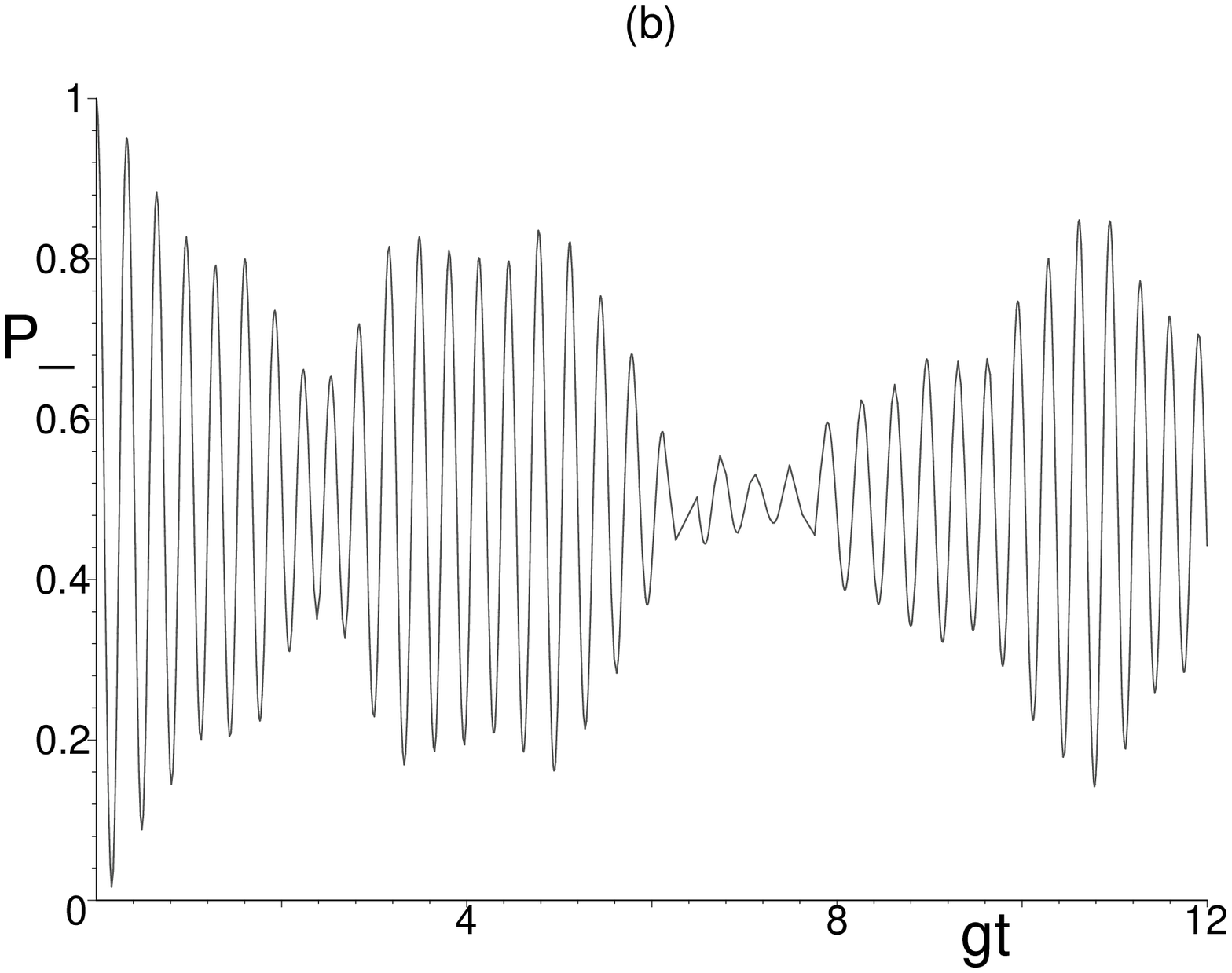}
  \caption{  Time evolution of the probability
  to find the ion in the internal ground state $P_-(gt)$ for $N=9$ (a) and $N=10$ (b)
respectively.}

\end{figure}
Exploiting an analytical method based on the analysis of the Rabi frequencies, it has been recently demonstrated
in Ref.~\cite{sab} that, starting from a total vibrational excitation number $N$, there exist an $N$-dependent
time instant at which the internal and external degrees of freedom of the trapped ion are disentangled ($c(t)=1$
or $c(t)=0$) or maximally entangled ($c(t)=1/2$). More in detail, it has been analytically proved that if $N \gg
1$ is odd,  at the  time instant $\bar{t}_o=\frac{\pi (N-1)}{4g}$ one has $c(\bar{t}_o)=1$ ($c(\bar{t}_o)=0$) if
$(N-1)/2$ is even (odd). This implies that  $S_v(\bar{t}_o)$ reaches its absolute minimum, indicating that the
internal and external degrees of freedom manifest a marked tendency to disentangle each other. On the contrary,
if $N \gg 1$ is even, at  the time instant $\bar{t}_e=\frac{\pi N}{4g}$, one has $c\left( \bar{t}_e \right)=1/2$
and therefore the vibrational entropy $S_v\left( \bar{t}_e \right)$ reaches its maximum value. This means that
the vibrational and electronic degrees of freedom are, at this time instant, maximally entangled. Fig.~2 displays
the time evolution of the probability $P_-(t) \equiv c(t)$ of finding the ion in its ground state (which can be
measured with the technique described at the end of subsection 2.2) according to Eq.~(\ref{ci}); Fig.~3 instead
shows the vibrational entropy $S_v (t)$, as given by Eq.~(\ref{entr2}). In both figures, the cases of an initial
total vibrational number $N=9$ (a) and $N=10$ (b) have been considered.

These figures illustrate in particular the
existence of a $N$-dependent time instant at which the system
under scrutiny exhibits different quantum behaviours dependent on
the parity of $N$. The interesting physical aspect is given by
the peculiar nonclassical sensitivity to the granularity of the initial total
number of vibrational quanta $N$. The physical origin of this
intrinsically quantum behaviour stems from the specific
two-boson coupling mechanism envisaged in this paper.
In the rest of the paper we shall analyse the robustness
of this quantum behaviour with respect to various experimental
imperfections as the fluctuations of the intensity of the coupling
lasers and a non-ideal initial state preparation

\begin{figure}
\centering
 \includegraphics[width=8cm]{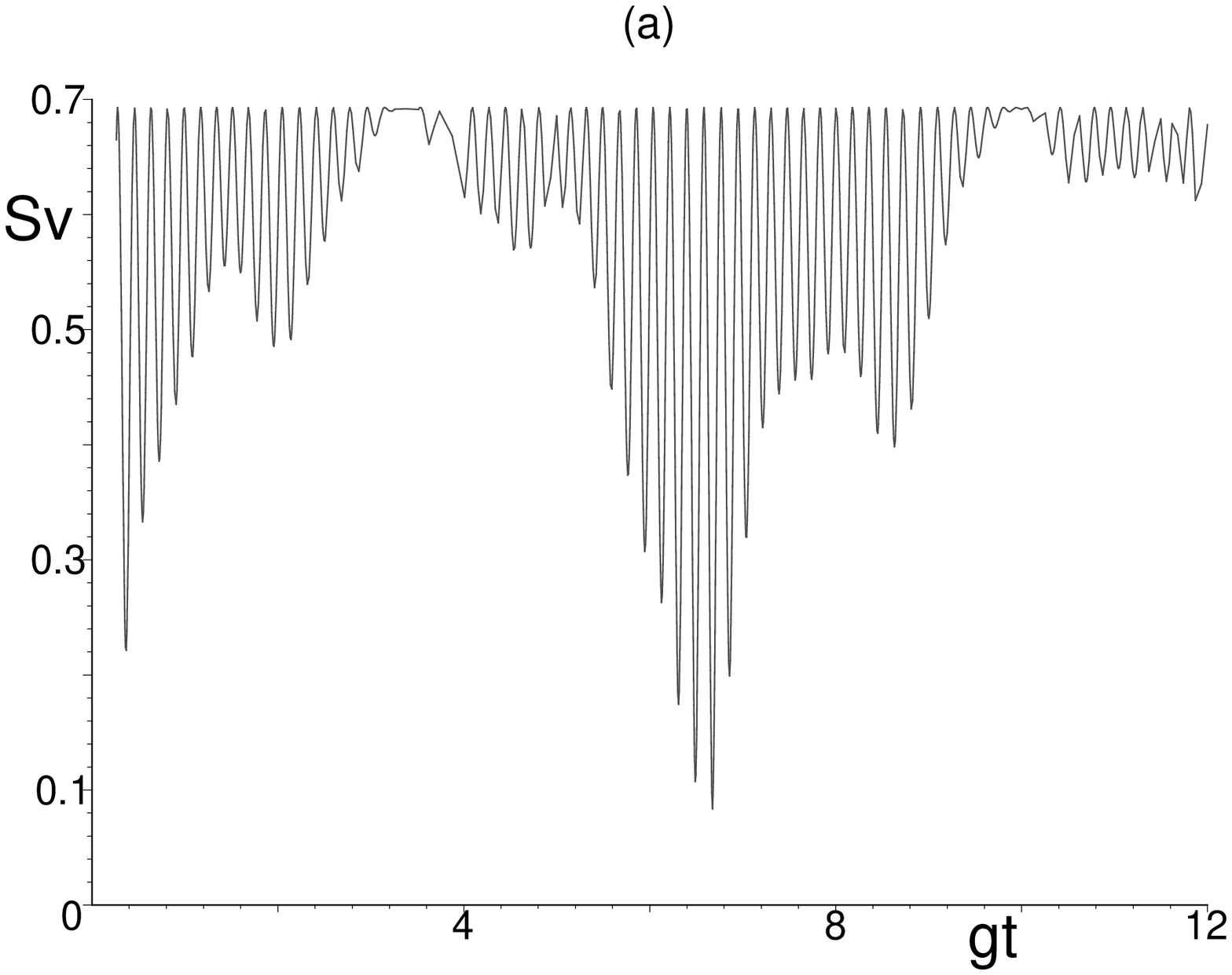}

\vspace{0.5cm}

 \includegraphics[width=8cm]{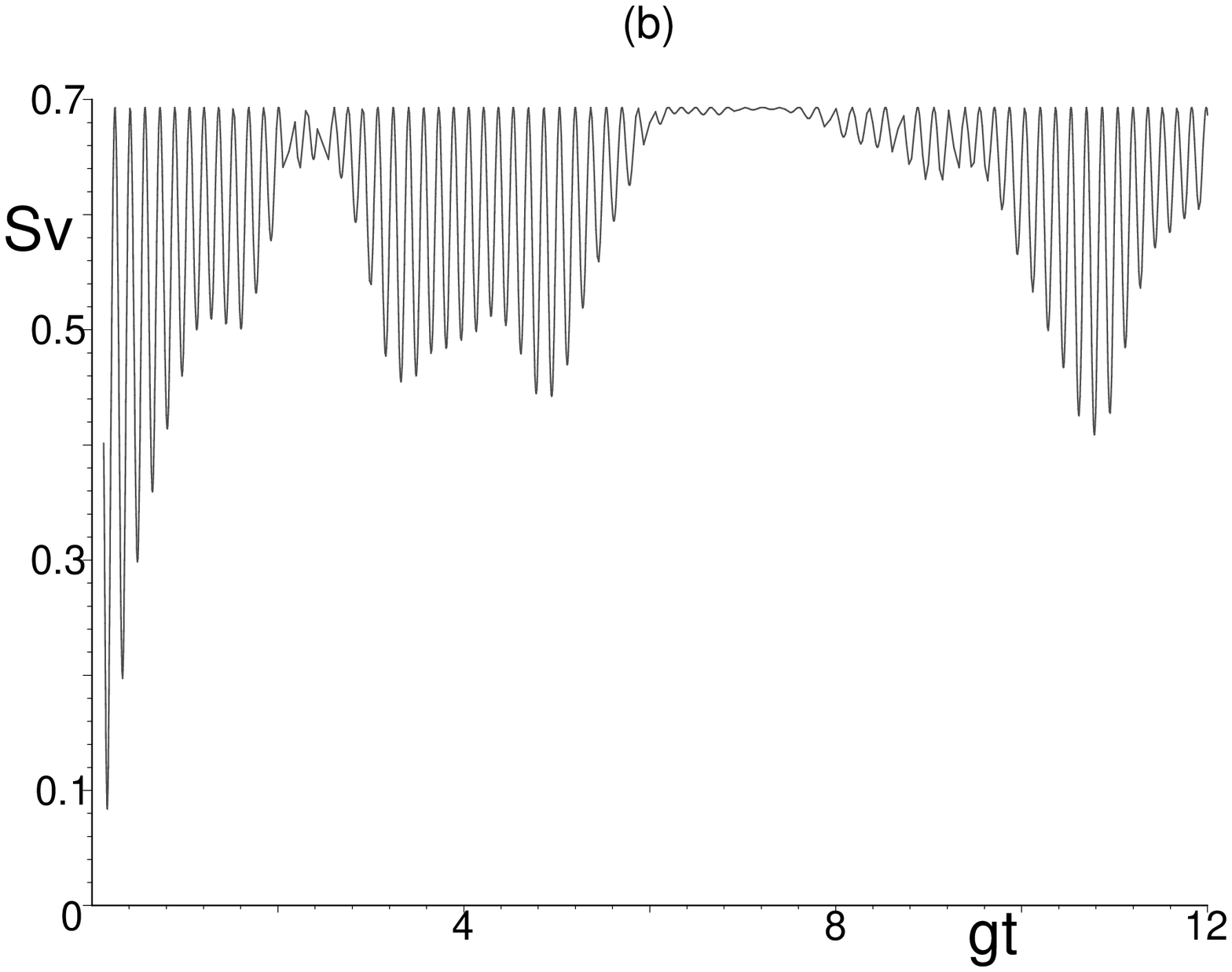}
  \caption{  Time evolution of the vibrational entropy
  $S_v(gt)$ for $N=9$ (a) and $N=10$ (b)
respectively.}

\end{figure}

\section{Influence of non-dissipative decoherence sources}

In the previous section we have assumed perfect unitary evolution
and initial state preparation for the system. This assumption is
partially justified by the fact that a very good control
of the quantum dynamics has been now achieved for trapped ions
and also because they are well isolated
from their environment. In fact, many experiments have shown that dissipative effects on the vibrational
motion can be considered completely negligible for long intervals of time. Moreover, spontaneous emission
is also practically irrelevant because the two internal states are
two hyperfine levels of the ground state.

However, decoherence effects have been nonetheless observed in the
motion of a laser driven trapped ion \cite{meek96,blatt}. In fact,
even when the entanglement with the environment is
negligible, fluctuations of some classical parameter
of the system may cause non-dissipative, phase-destroying
effects.  In the trapped ion case, phase
decoherence is mainly caused by the fluctuations of the Rabi frequency,
which are
induced by the laser intensity fluctuations, making the coupling
constant $g$ fluctuate.
A quantitative explanation of the decoherence observed in
the Rabi oscillations of Ref.~\cite{meek96}, in terms of a fluctuating
laser pulse area, has been recently provided by Bonifacio {\it et al}.
in Ref.~\cite{bonifacio}, using a model-independent formalism
able to describe non-dissipative decoherence phenomena due to the
fluctuations of classical parameters or internal variables of a system.

In the model-independent approach of Ref.~\cite{bonifacio},
the dynamical quantities in the presence of non-dissipative
decoherence can be obtained simply through an average over
an appropriate probability distribution of the fluctuating
parameter. In the trapped ion case, the random parameter is the
positive dimensionless random variable $A(t)=\int_0 ^t d\xi
g(\xi)$,
which is proportional to the laser pulse area. In fact,
the quantity $g(\xi)$ is a stochastic coupling constant, whose
fluctuations have to be traced back to the laser intensity fluctuations.
The time evolution of a generic dynamical quantity $O(t)$
becomes therefore the averaged quantity
\begin{equation}
\bar{O}(t)= \int_{0}^{\infty} dA P(t,A) O(A) \;.
\label{robara}
\end{equation}
The probability distribution $P(t,A)$ is the Gamma distribution
function
\begin{equation}
P(t,A) =  \frac{e^{-A/g\tau}}{g\tau}
\frac{(A/g\tau)^{(t/\tau)-1}}
{\Gamma(t/\tau)} ,
\label{agamma}
\end{equation}
obtained in Ref.~\cite{bonifacio} by imposing the semigroup condition
for the averaged density matrix of the system.
The physical meaning of the parameters $g$ and $\tau$ in Eq.~(\ref{agamma})
can be easily understood by considering the mean and the variance of
the probability distribution (\ref{agamma}),
\begin{eqnarray}
&& \langle A \rangle =g t \\
&& \sigma^2(A)=\langle A^2\rangle - \langle A \rangle ^2 = g^2 t\tau,
\end{eqnarray}
implying that $g$ has now to be meant as a {\em mean}
coupling constant, and that $\tau $ quantifies the strength of $A$ fluctuations.
In Ref.~\cite{bonifacio} the decay of the Rabi oscillations observed
in Ref.~\cite{meek96} is well fitted assuming
$\tau \simeq 1.5 \cdot 10^{-8}$ sec (see also \cite{milburn}). In this case,
the relevant experimental timescale
$t$ is much larger than $\tau$ and in this limit the Gamma
distribution function can be well approximated by the following
Gaussian distribution
\begin{equation}
P(t,A) \simeq \frac{1}{\sqrt{2\pi g^2 t \tau}}
\exp\left\{-\frac{(A-g t)^2}{2g^2 t \tau}
\right\}.
\label{gauap}
\end{equation}
\begin{figure}
\centering
 \includegraphics[width=8cm]{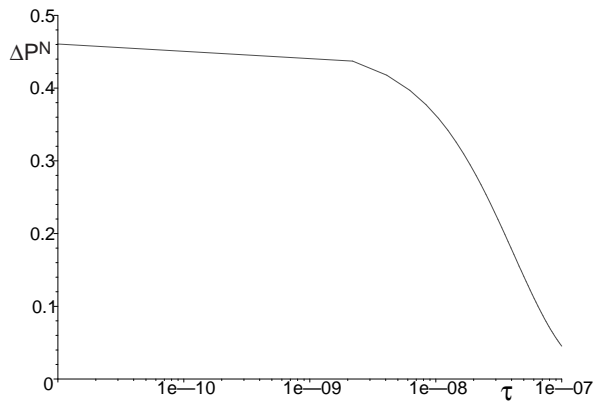}
  \caption{  The probability
  difference $\Delta P^{N}$ defined in the text,
  at the time instant $\bar{t}$, as a
function of the laser pulse area fluctuation strength parameter $\tau$ (expressed in seconds), in the case of a
total initial vibrational quanta $N=9$ and for $g=10^{5}$ Hz.}
 \end{figure}
The experimentally observed quantity is the probability of finding the
ion in its ground electronic state $P_-(t)\equiv c(t)$.  In the
presence of the fluctuating coupling constant $g(\xi)$ caused
by the intensity fluctuations of the coupling lasers, this probability
has therefore to be averaged over the Gaussian distribution
(\ref{gauap}). Using Eq.~(\ref{ci}) we find
\begin{equation}
\bar{P}_-(gt)=\frac{1}{2} \left[ 1+ \sum_{k=0}^{N} e^{-8 g^2 \tau
(N-k)k t} \cos(4 \sqrt{(N-k)k}gt)   \right] , \label{pdeco}
\end{equation}
where the quantum coherent oscillations at frequencies $f_{k}$ are
now exponentially decaying because of the non-dissipative
decoherence.

To quantify the robustness of the parity-dependent entanglement
with respect to the various experimental imperfections,
we have considered the quantity $\Delta P^{N}$, defined as the
difference between the values assumed by $\bar{P}_-(g \bar{t})$ in
correspondence to $N$ (odd) and $N+1$ total initial
vibrational excitations, at the ``intermediate''
time $\bar{t}= \frac{\bar{t}_e +
\bar{t}_o}{2}$. With no imperfections and $N \gg 1$, one has
$\bar{t} \sim \bar{t}_e \sim \bar{t}_o $, and
$\bar{P}_-^{N=odd}(g \bar{t}) \sim 1$, $\bar{P}_-^{N+1}(g \bar{t}) \sim
1/2$, so that $\Delta P^{N} \simeq 1/2$ whatever $N$ is. We emphasize that such a result highlights the fact that the probability $P_-(\bar{t})$ of finding the ion in its electronic ground level at $t=\bar{t}$ depends on $N$ being even or odd. As long as the effect of phase
decoherence increases, the difference between the odd $N$ case and
the even $N$ case tends to zero, because in both cases the system state
becomes the same statistical mixture and $\Delta P^{N}$ tends to
zero.

In Fig.~4 we show the dependence of the probability difference $\Delta P^{N}$ as a function of the laser pulse
area fluctuation strength $\tau$, in the case when $N=9$ and $g = 10^5 Hz$. One can see the expected decay to
zero of $\Delta P^{N}$ for increasing $\tau$ (i.e., increasing laser intensity fluctuations); in particular one
has a well visible transition from a quantum behaviour at small $\tau$ to a completely decoherent behaviour at
larger $\tau$, taking place at a threshold value $\tau_{th} \sim 3 \cdot 10^{-8}$ sec. Since current experimental
situations correspond to $\tau \leq 10^{-8}$ sec, this means that the parity-dependent entanglement effect
described in the preceding section can be experimentally detected even with the unavoidable presence of
non-dissipative decoherence sources, as for example laser intensity fluctuations.

\section{Imprecision in the initial state preparation}

Another important condition for the observability of the parity
effect reported in this paper is the possibility of
preparing the initial vibrational Fock state $|N\rangle $
along the direction $X'$ at $\pi/4$ radians with respect to $X$.
It is however not difficult to guess that at the end of any scheme
aimed at preparing a vibrational ionic Fock state containing exactly
$N$ vibrational quanta, the state effectively reached is a
vibrational density matrix
$\rho_{v}(0)$ containing also diagonal terms
diffeent from $\vert N \rangle \langle N \vert$ in the total vibrational
energy eigenbasis. Of course, such a mixture of initial odd and even
total vibrational quanta counters the possibility of observing the parity
dependent entanglement. However, we shall show that the degree of
preparation efficiency required in order to keep the parity effect in the
dynamics of our system appears realistically in the grasp of the
experimentalists. In other words we shall prove that the occurrence
of parity effects, although attenuated, turns out to be compatible
with the initial state preparation fidelity presently achievable
for ionic vibrational states.

We assume that at $t=0$ the initial vibrational
state can be described as follows:
\begin{equation}
\rho_{v} (0) = \sum_{m=0}^{\infty} {\cal N} e^{- (m-N)^2 / 2 \Delta^2}
\vert m,0 \rangle \langle m, 0 \vert,  \label{ro}
\end{equation}
${\cal N}$ being the normalization constant.
Equation\ (\ref{ro}) is a gaussian weighted mixture of pure Fock states
$\vert m,0 \rangle \langle m,0 \vert $ along the direction
$X'$, centered on a prefixed number
$N$ of vibrational quanta and controlled by a width parameter
$\Delta$. It is immediate to see that when $\Delta \rightarrow 0$
$\rho_{v}(0)$ tends to $\vert N,0 \rangle \langle N,0 \vert$. The form
given to $\rho_{v}(0)$ suggests the introduction of the parameter
\begin{equation}
\eta = 1- \frac{p_{N+1}}{p_N} =1-e^{-1/2\Delta^{2}} \label{eta}
\end{equation}
describing the efficiency of initial state preparation. $p_{N+1}$ and
$p_N$ appearing in Eq.~(\ref{eta}) are the probabilities of
having $N+1$ or $N$ quanta in the initial vibrational Fock state.
With the help of equations (\ref{pdeco}) and (\ref{ro}), it is not
difficult to verify that, in the presence
of both laser intensity fluctuations and imperfect initial
state preparation, the probability to find the ion in the internal
ground state $\vert -\rangle $,
$\bar{P}_-(gt)$, assumes the form
\begin{equation}
\bar{P}_-(gt)=\frac{1}{2} \left[ 1+ \sum_{m=0}^{\infty} {\cal N} e^{-
(m-N)^2 / 2 \Delta^2} \sum_{k=0}^{m} e^{-8 g^2 \tau (m-k)k t}
\cos(4\sqrt{(m-k)k}gt)   \right] \label{pdeco2}
\end{equation}
As we have done in the preceding section, we quantify the robustness of the parity-dependent effect in terms of
the probability difference $\Delta P^{N}$ defined above. In Fig.~5 $\Delta P^{N}$ is plotted as a function of the
preparation efficiency $\eta$  for three different values of the fluctuation strength parameter, $\tau = 10^{-9}$
sec (solid line), $\tau = 10^{-8} $ sec (dotted line) and $\tau = 10^{-7} $ sec (dashed line), again in the case
$N=9$, $g=10^{5}$ Hz. As expected, one has a steady increase of $\Delta P^{N}$ for increasing preparation
efficiency. In particular, if we assume the realistic parameters $\eta=0.9$ and $\tau =10^{-8}$ sec, one can see
that the parity dependent entanglement effect is still visible, although attenuated by almost 40\%.

\begin{figure}
\centering
 \includegraphics[width=8 cm]{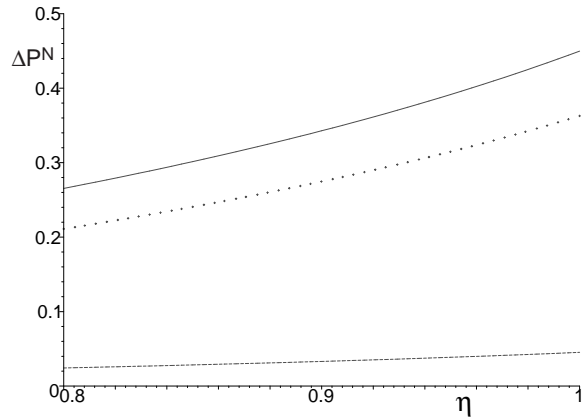}
  \caption{ The probability difference $\Delta P^{N}$
  at the time instant $\bar{t}$ as a function of the
efficiency of the initial state preparation $\eta$ (see Eq.~(\protect\ref{eta})) for $\tau = 10^{-9}$ sec (solid
line), $\tau = 10^{-8}$ sec (dotted line) and $\tau = 10^{-7} $ sec (dashed line). The other parameters are as in
Fig.~3. }

\end{figure}

\section{Conclusions}

In this paper we have analyzed how some experimental imprecisions 
influence a nonclassical effect we have brought to light and discussed 
in a previous paper \cite{sab}, namely the parity dependent entanglement effect.
According to this effect by changing only one vibrational quantum in 
an initial condition characterized by a total number of oscillatory 
quanta $N \gg 1$, the state of the system, at approximately the 
same instant of time $t_e \simeq t_o$, drastically changes.
Indeed at this time instant the vibrational and electronic degrees of 
freedom of the ion are maximally entangled  
(i.e. maximum reduced entropy) for even $N$ and disentangled for odd $N$.

The parity dependent quantum effect has a nonclassical origin since it 
stems from the granularity of the oscillatory energy of the center of 
mass of the ion. For this reason we believe it's worth being 
experimentally verified.
We have thus studied the robustness of the
effect with respect to various experimental imperfection. Since single
trapped ions are essentially dissipationless systems, we have
considered the effect of non-dissipative decoherence sources as the intensity
fluctuations of the coupling lasers, and also the effect of an
imperfect initial state preparation.
The main result of the paper is that the parity-dependent entanglement
is quite robust against these experimental imperfections and that
it can be experimentally seen using presently available technology.

\section*{Acknowledgements}

The authors of Palermo University acknowledge financial support from
CRRNSM-Regione Sicilia.  One of the authors (A.N.) is indebted to
MURST and FSE for supporting this work through Assegno di Ricerca.

\section*{References}

\end{document}